\documentclass{pazhb_eng}
\usepackage[utf8]{inputenc}
\usepackage{graphicx}
\usepackage{natbib}
\usepackage{color}

\usepackage{multirow}

\definecolor{darkblue}{rgb}{0,0,0.9}

\def\arcsec{$^{\prime\prime\,}$}
\def\arcmin{$^{\prime\,}$}

\def\ergs{erg s$^{-1}$}
\def\flux{erg cm$^{-2}$ s$^{-1}$}
\def\deg{\hbox{$^\circ$}}

\def\AA{\buildrel _{\hskip 0.5pt \circ} \over {\mathrm{A}}}

\begin{document}

\journalinfo{2013}{39}{8}{513}[522]

\title{Identification of Four X-ray Sources from the INTEGRAL and Swift Catalogs}

\author{A.\,A.\,Lutovinov\email{aal@iki.rssi.ru}\address{1}, A.\,I.\,Mironov\address{1,2}, R.\,A.\,Burenin\address{1},
M.\,G.\,Revnivtsev\address{1}, S.\,S.\,Tsygankov\address{3,4,1}, M.\,N.\,Pavlinsky\address{1},
I.\,V.\,Korobtsev\address{5} and M.\,V.\,Eselevich\address{5}\\
\bigskip
  {\it (1) Space Research Institute, Russian Academy of Sciences, Moscow, Russia}\\
  {\it (2) Moscow Institute of Physics and Technology, Dolgoprudnyi, Moscow reg., Russia}\\
  {\it (3) Finnish Centre for Astronomy with ESO (FINCA), University of Turku, Piikki\"o, Finland}\\
  {\it (4) Astronomy Division, Department of Physics, University of Oulu, Finland}\\
  {\it (5) Institute for Solar-Terrestrial Physics, Russian Academy of Sciences, Siberian Branch, Irkutsk, 664033 Russia}
}

\shortauthor{Lutovinov et al.}

\shorttitle{Identification of four X-ray sources}

\submitted{18 March 2013}

\begin{abstract}

Four hard X-ray sources from the INTEGRAL and Swift catalogs have been identified. X-ray
and optical spectra have been obtained for each of the objects being studied by using data from the
INTEGRAL, Swift, ROSAT, and Chandra X-ray observatories as well as observations with the RTT-150
and AZT-33IK optical telescopes. Two sources (SWIFT J1553.6+2606 and SWIFT J1852.2+8424) are
shown to be extragalactic in nature: the first is a quasar, while the registered X-ray flux from the second is the
total emission from two Seyfert 1 galaxies at redshifts 0.1828 and 0.2249. The source IGR J22534+6243
resides in our Galaxy and is an X-ray pulsar with a period of $\sim46.674$ s that is a member of a high-mass
X-ray binary, probably with a Be star. The nature of yet another Galactic source, SWIFT J1852.8+3002, is not
completely clear and infrared spectroscopy is needed to establish it.

\englishkeywords{X-ray sources, active galactic nuclei, X-ray binaries}

\end{abstract}

\section{Introduction}
\label{sec:intro}

The value of astrophysical sky surveys in various
wavelength ranges for their subsequent use in
studying the physical and statistical properties of
various populations of sources is directly related both
to the completeness of the surveys themselves and to
the completeness of identifying and determining the
nature of the objects detected in them. At present, the
most complete all-sky hard X-ray ($>15$ keV) surveys
are the INTEGRAL (Krivonos et al. 2010a, 2012;
Bird et al. 2010) and Swift (Cusumano et al. 2010;
Baumgartner et al. 2013) surveys. These have
a very high identification completeness of the detected
sources; in particular, this completeness
reaches 92\% in the INTEGRAL Galactic survey
(Krivonos et al. 2012). Such a high percentage has
been reached through the long-term work of several
scientific groups in the world (see, e.g., the review
by Parisi et al. 2013, and references therein; Masetti
et al. 2007, 2010; Tomsick et al. 2008, 2009), including
our work (Bikmaev et al. 2006, 2008; Burenin et al. 2008, 2009; Lutovinov et al. 2012a, 2012b;
Karasev et al. 2012), using soft X-ray ($<10$ keV),
optical, and infrared observations.

This work is the next one in our program on
the localization and identification of hard X-ray
sources from the INTEGRAL and Swift catalogs.
Four objects from the INTEGRAL Galactic survey
catalog (Krivonos et al. 2012) and the Swift 70-
month all-sky catalog (Baumgartner et al. 2013)
were included in the sample: IGR J22534+6243,
SWIFT J1553.6+2606, SWIFT J1852.2+8424, and
SWIFT J1852.8+3002. Apart from the results of our
optical observations, we also present the results of
our spectral and timing analysis for these sources
obtained from INTEGRAL, Swift, ROSAT and Chandra data.

\section{Observations and data analysis}
\label{sec:data}

The objects being investigated here were detected
by the IBIS/INTEGRAL (Winkler et al. 2003) and
BAT/Swift (Gehrels et al. 2004) telescopes as weak
hard X-ray ($>15$ keV) sources. The typical positional
accuracy of such objects for the above instruments
is $4-7$\arcmin\ (Krivonos et al. 2010b; Tueller et al. 2010),
which makes their identification in the optical and
infrared wavelength ranges virtually impossible. The
sky regions around SWIFT J1553.6+2606,
SWIFT J1852.2+8424, and SWIFT J1852.8+3002
were observed by the XRT/Swift telescope, which
allowed one to detect them in the soft X-ray ($0.6-10$ keV) energy band and to improve the positional
accuracy to a few arcseconds (Baumgartner
et al. 2013). IGR J22534+6243 was first detected
on the total sky map obtained during the
INTEGRAL nine-year Galactic survey (Krivonos
et al. 2012). The study of the archival data showed
that the sky region around IGR J22534+6243 was
previously observed by the ROSAT and Chandra
observatories as well as by the XRT/Swift telescope
and that a soft X-ray source that can be
identified with the objects 1RXS J22535.2+624354,
CXOU J225355.1+624336, and 2MASS J22535512+6243368 (Landi et al. 2012; Israel
and Rodriguez 2012) is registered at a statistically
significant level within the INTEGRAL error circle in
all these observations. In addition, based on Chandra
and XRT/Swift data, Halpern (2012) detected X-ray
pulsations from this object with a period of $\sim46.67$ s.
In combination with the properties of the optical star
(Masetti et al. 2012), this allowed IGR J22534+6243
to be presumably classified as belonging to the class
of X-ray pulsars that are members of high-mass Xray
binaries. The coordinates of all four objects being
studied that we obtained by analyzing the XRT data
are given in Table 1; the positional accuracy is $\simeq3.5$\arcsec.

\begin{table}[]
\centering

\footnotesize{
   \caption{List and coordinates of sources}
\medskip
   \begin{tabular}{lcc}
     \hline
     \hline

          Name &  RA     & Dec     \\
              & (J2000) & (J2000)      \\[1mm]
     \hline
     SWIFT\,J1553.6+2606 &  15$^h$ 53$^m$ 34.91$^s$ &  26\deg 14\arcmin 41.8\arcsec \\
     SWIFT\,J1852.2+8424A&  18$^h$ 50$^m$ 24.76$^s$ &  84\deg 22\arcmin 41.0\arcsec \\
     SWIFT\,J1852.2+8424B&  18$^h$ 46$^m$ 50.25$^s$ &  84\deg 25\arcmin 02.2\arcsec \\
     SWIFT\,J1852.8+3002 &  18$^h$ 52$^m$ 49.50$^s$ &  30\deg 04\arcmin 27.3\arcsec \\
     IGR\,J22534+6243    &  22$^h$ 53$^m$ 55.10$^s$ &  62\deg 43\arcmin 36.8\arcsec \\
     \hline
    \end{tabular}
    }
\end{table}

The main optical observations were carried out on
the night of July 5, 2012, with the 1.5-m Russian-Turkish (RTT-150) telescope using the \emph{TFOSC}\footnote{http://hea.iki.rssi.ru/rtt150/ru/index.php?page=tfosc}
medium- and low-resolution spectrograph. For our
spectroscopy, we used grism N15 with a spectral resolution
of $\approx12\AA$ (the full width at half maximum that
gives the widest wavelength range ($3500 - 9000\AA$)
and the highest quantum efficiency; the signal integration
time was 900 s for SWIFT J1852.8+3002
and IGR J22534+6243, 1200 s for
SWIFT J1852.2+8424A, and 1800 s for
SWIFT J1553.6+2606 and SWIFT J1852.2+8424B.
Additional optical spectroscopy for
IGR J22534+6243 was performed on March 1, 2013,
with the AZT-33IK telescope at the Sayansk Observatory
of the Institute for Solar-Terrestrial Physics,
the Siberian Branch of the Russian Academy of
Sciences. For these observations, we used the UAGS
spectrograph mounted at the Cassegrain focus of
the telescope and equipped with a $1300$ lines/mm
grating, which made it possible to take a spectrum
in the range $6250-6850\AA$, near the $H\alpha$ line, with a
resolution of about $4\AA$.

We processed the optical data in a standard way,
using the \emph{IRAF}\footnote{http://iraf.noao.edu} software and our own software.

For the spectral and timing analysis of the sources
in the $0.2-10$ keV energy band, we used XRT/Swift
and ROSAT and Chandra (for IGR J22534+6243) observational
data. They were processed with the appropriate
software\footnote{http://swift.gsfc.nasa.gov and http://cxc.harvard.edu/ciao/} and the FTOOLS 6.11 software package.
A hard X-ray ($>20$ keV) spectrum of
IGR J22534+6243 was reconstructed from INTEGRAL data using
software developed at the Space Research Institute of
the Russian Academy of Sciences (for more details, see Krivonos et al. 2010b).

\section{Results}
\label{sec:res}

Most of the objects from Table 1 lie fairly high
above the Galactic plane, while IGR J22534+6243,
though this source is close to it ($b\simeq3$\deg), is nevertheless
far from the central Galactic regions ($l\simeq110$\deg). Therefore, in principle, the positional accuracy
($\simeq3.5$\arcsec) is high enough for soft X-ray sources to be
identified at optical wavelengths. On the other hand,
the weak objects from the Swift catalog detected
by the BAT telescope have a positional accuracy of
$\sim6-7$\arcmin\ (Tueller et al. 2010). This can lead to ambiguities
even at the stage of their identification with soft
X-ray sources. There are two such sources among
the four objects from Table 1.

\bigskip

\subsection*{SWIFT\,J1553.6+2606}

\begin{figure*}
\vbox{
\hbox{
\includegraphics[width=\columnwidth,bb=37 169 574 621,clip]{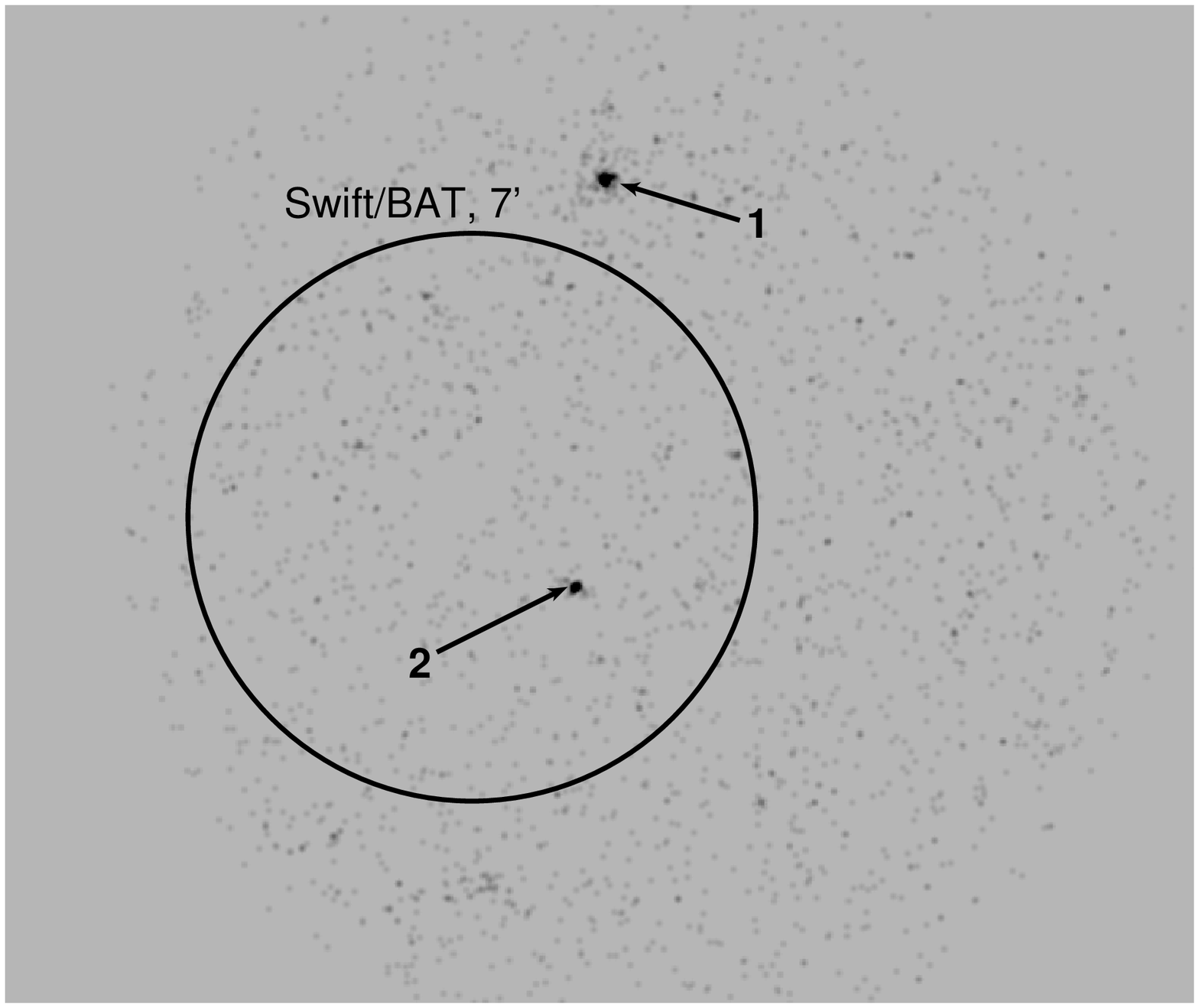}
\hspace{3mm}\includegraphics[width=\columnwidth,bb=37 169 574 621,clip]{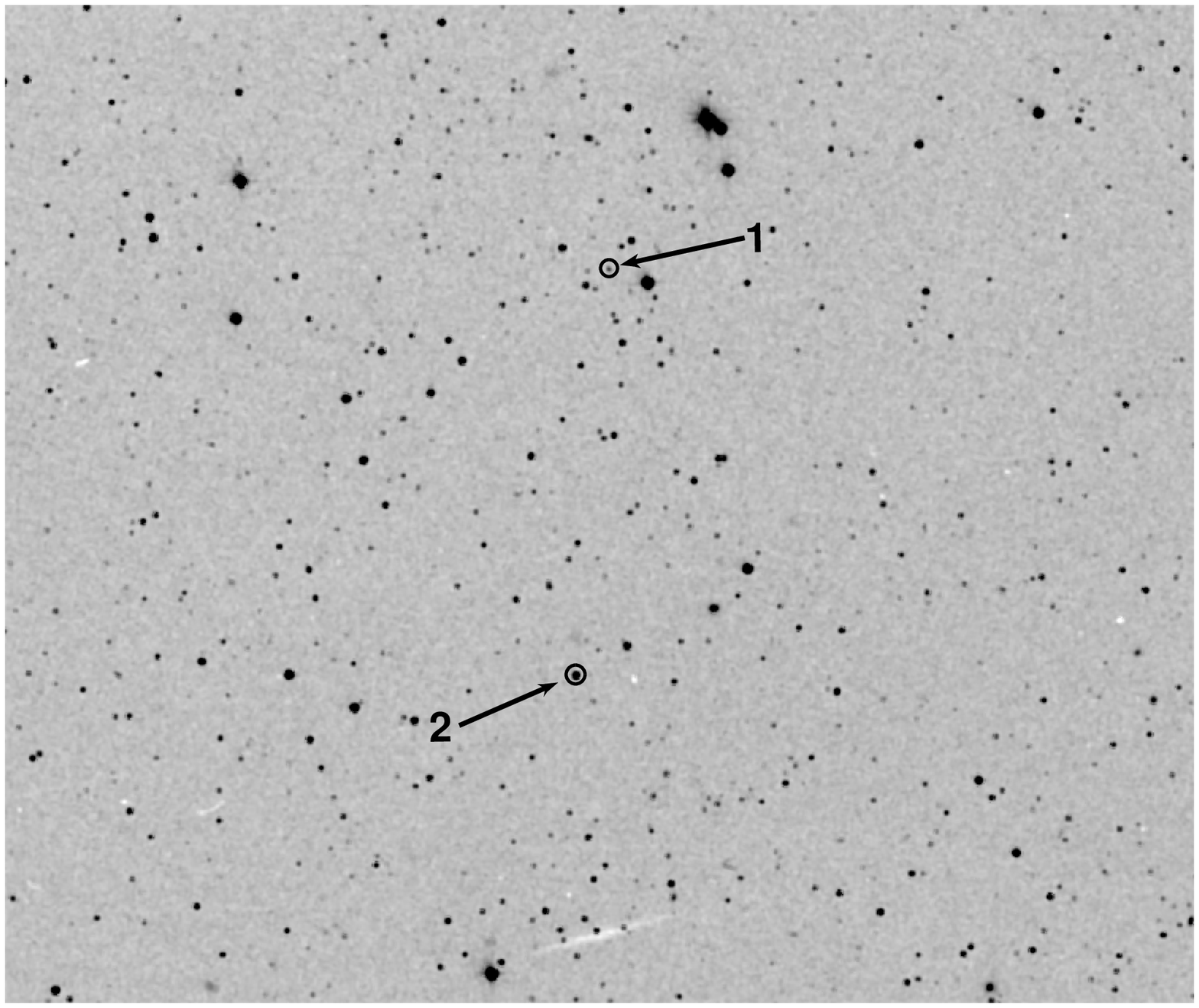}
}
\hbox{
\includegraphics[width=1.02\columnwidth,bb=33 167 568 670,clip]{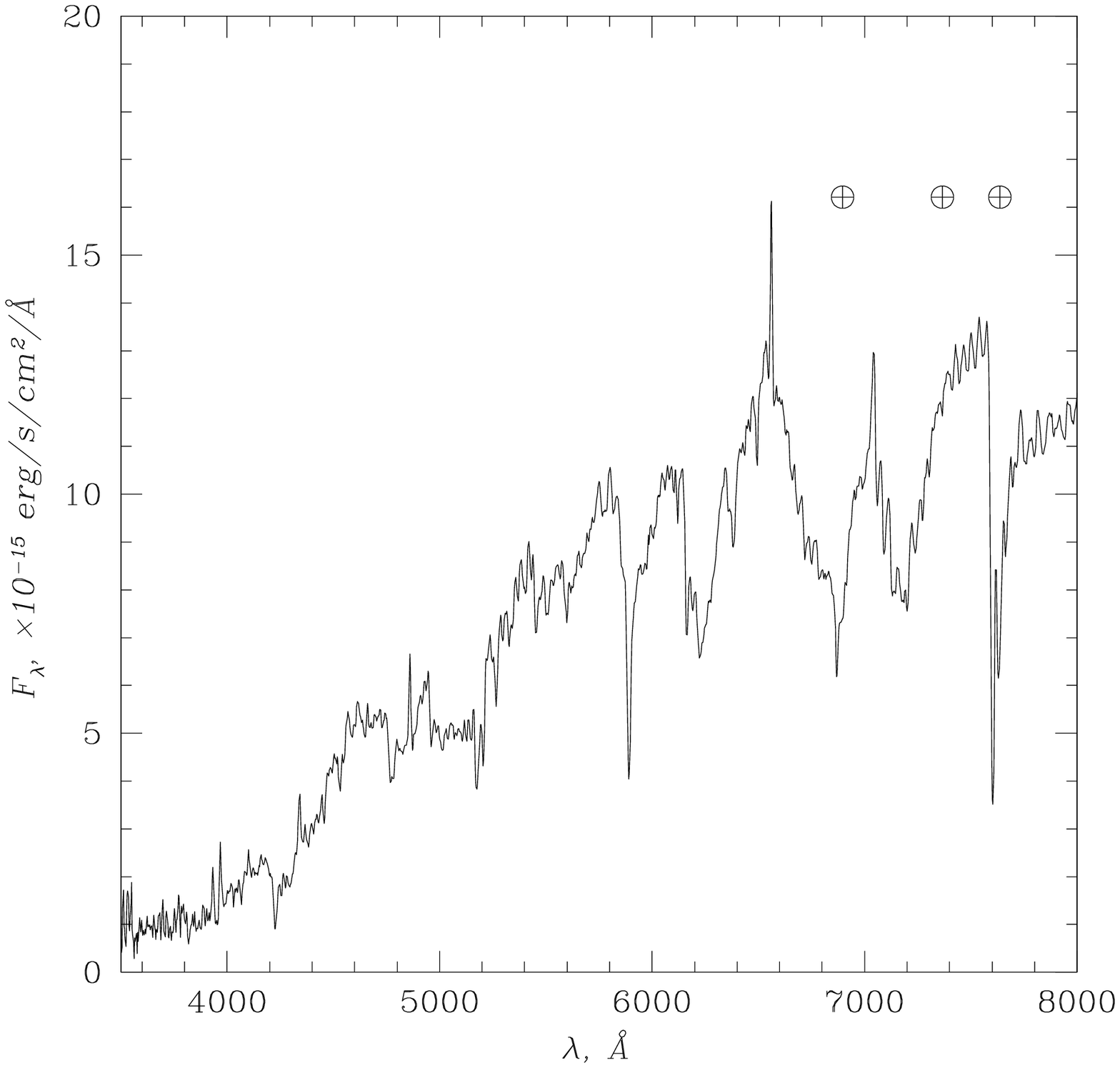}
\includegraphics[width=0.98\columnwidth,bb=140 270 547 680,clip]{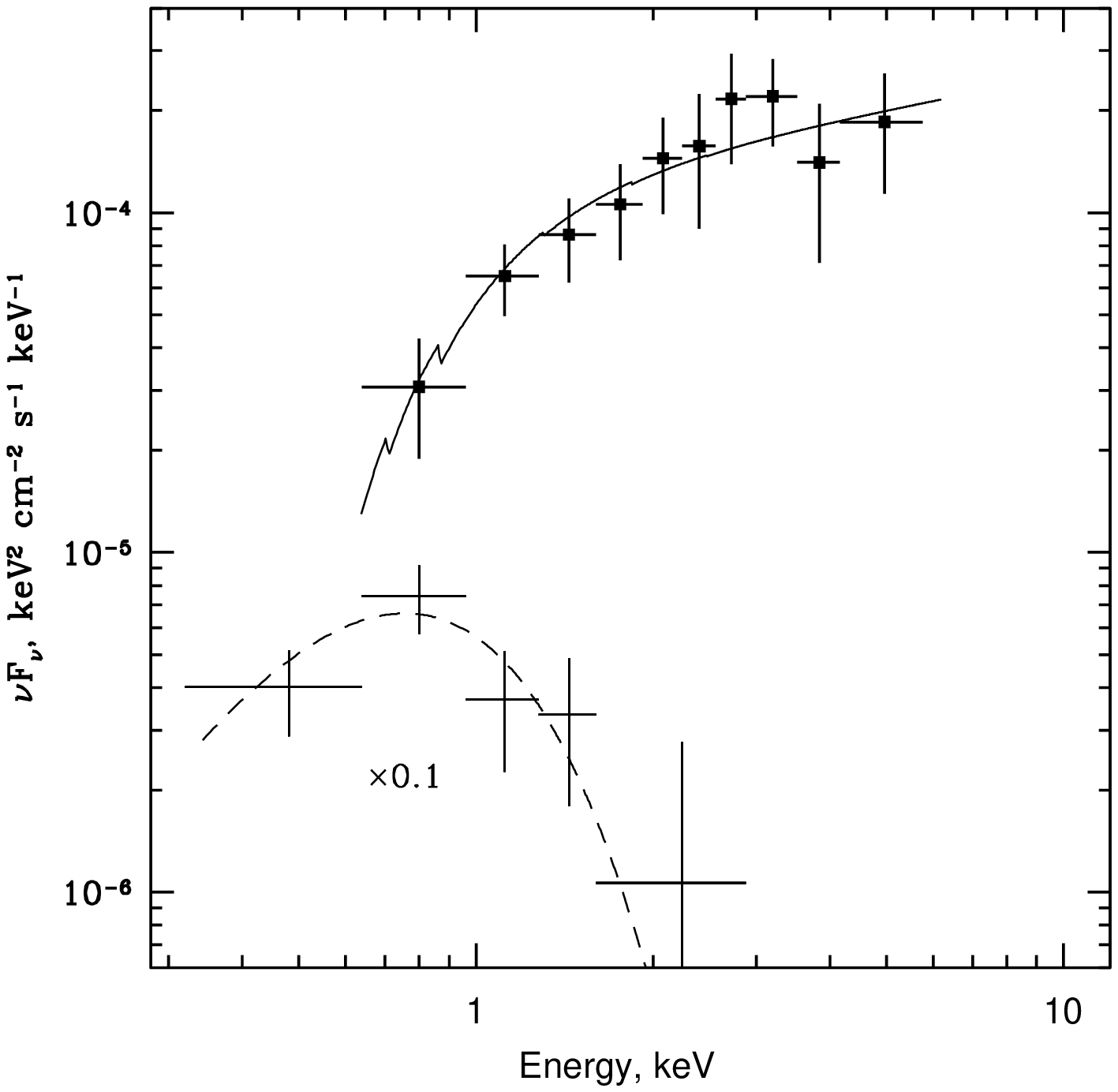}
}
}
\caption{(a) X-ray and (b) optical images of the sky regions around SWIFT J1553.6+2606. The circle indicates the BAT error
circle of the object (7\arcmin\ in radius). Numbers 1 and 2 and the arrows mark the positions of its presumed X-ray and optical
counterparts. (c) The RTT-150 optical spectrum of the second source. (d) The XRT energy spectra of sources 1 (dots) and 2
(crosses). The solid and dashed lines, respectively, indicate the best fits. For clarity, the spectrum of source 2 was multiplied
by 0.1. \label{swiftj15536}}
\end{figure*}
\begin{figure*}
\vbox{
\hbox{
\includegraphics[width=\columnwidth,bb=37 169 574 621,clip]{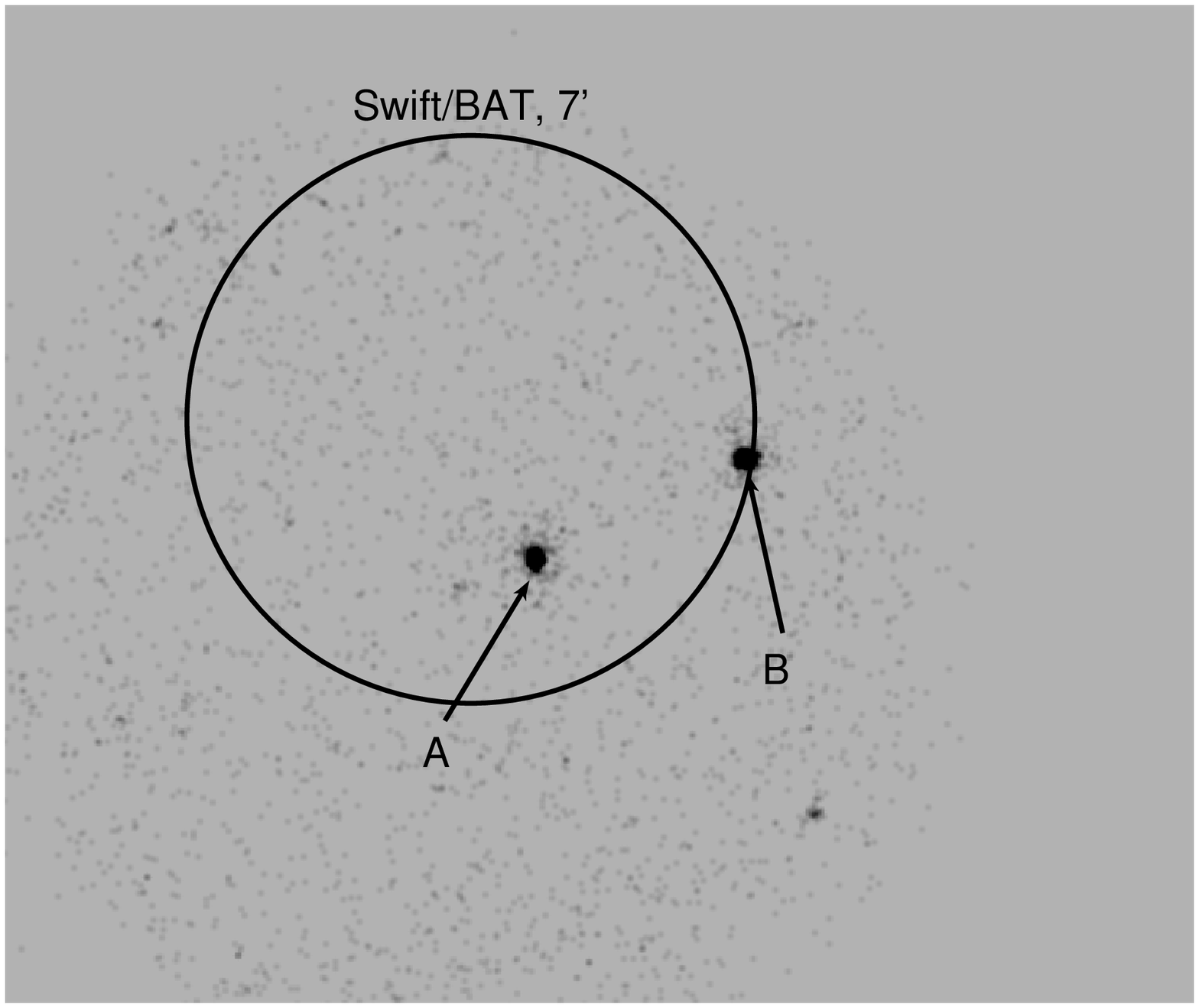}
\hspace{3mm}\includegraphics[width=\columnwidth,bb=37 169 574 621,clip]{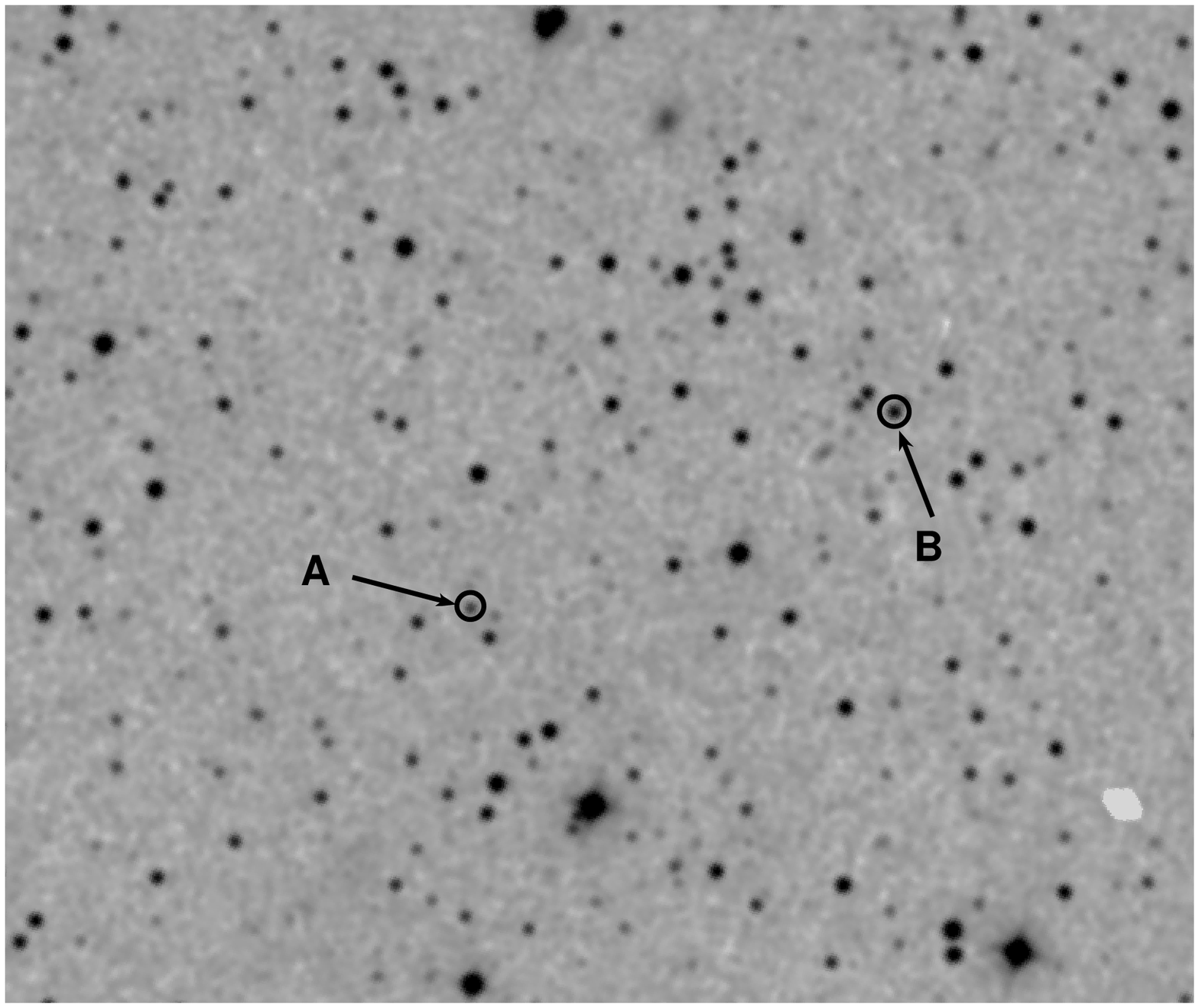}
}
\hbox{
\includegraphics[width=\columnwidth,bb=33 167 568 681,clip]{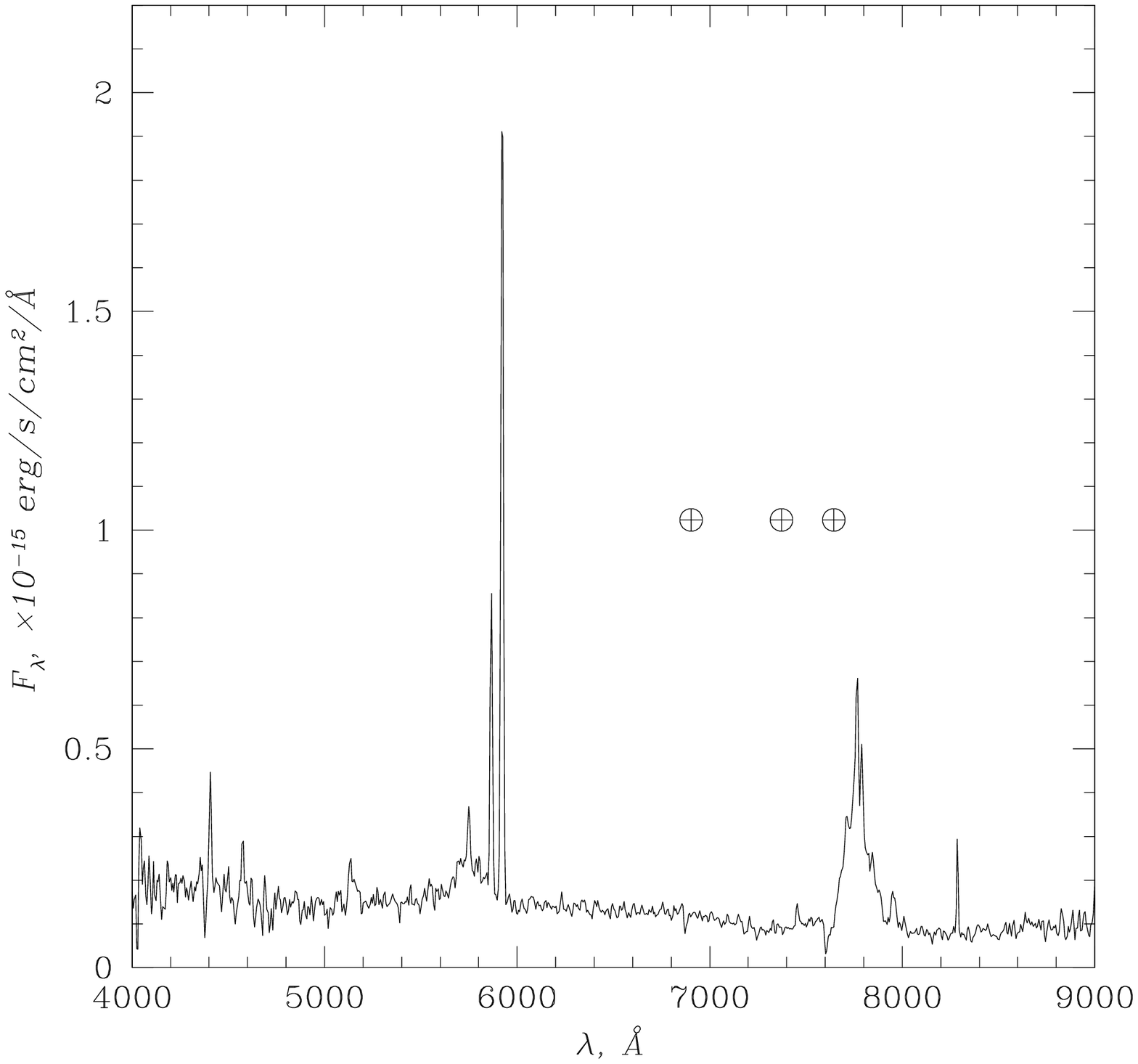}
\includegraphics[width=\columnwidth,bb=33 167 568 681,clip]{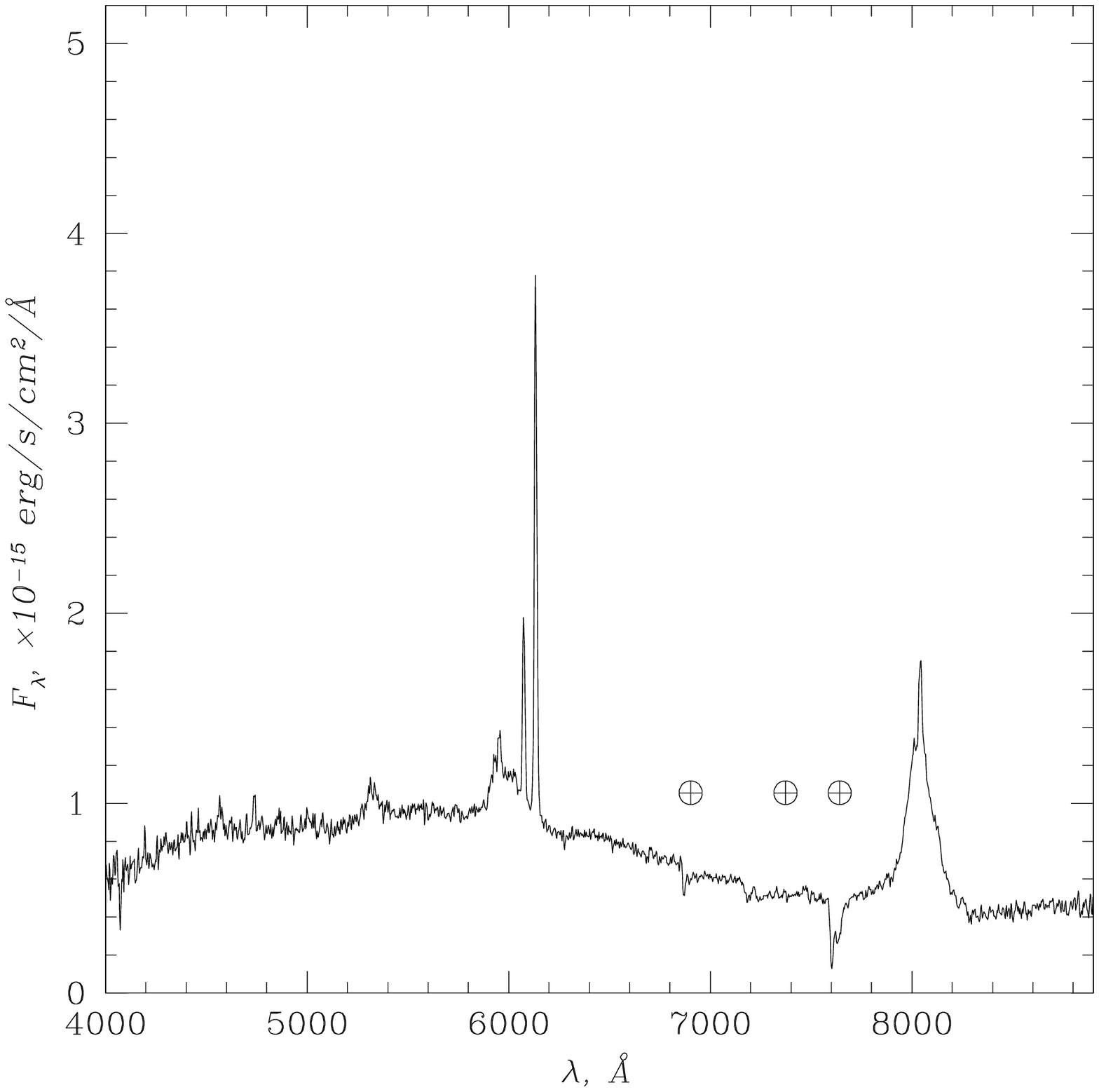}
}
}
\caption{(a) Soft X-ray and (b) optical images of the sky regions around SWIFT J1852.2+8424. The circle indicates the BAT
error circle (7\arcmin). Letters A and B indicate the positions of its X-ray and optical counterparts. The RTT-150 optical spectra of
sources A (c) and B (d). \label{swiftj18522a}}
\end{figure*}

In the Swift 70-month catalog (Baumgartner et al.
2013), the object SDSS J155334.73+261441.4,
which is a quasar at redshift $z=0.1664\pm0.0013$, is
specified as an optical counterpart of SWIFT J1553.6+2606. However, the separation
between the BAT position of SWIFT J1553.6+2606
and the quasar's position in the sky is $\simeq9$\arcmin, which exceeds
the BAT positional accuracy (Tueller et al. 2010).
At the same time, within the Swift/BAT error circle
there is another X-ray source with coordinates RA=15$^h$ 53$^m$ 38.12$^s$, Dec=26\deg 04\arcmin 38.8\arcsec\ that can also
be associated with SWIFT J1553.6+2606. This fact
is illustrated in Fig. 1a, where an XRT X-ray image
of the sky region around SWIFT J1553.6+2606
is shown; numbers 1 and 2 indicate the quasar
SDSS J155334.73+261441.4 and the second soft Xray
source.

Figure 1b shows the same sky region at optical
wavelengths based on red Palomar Digital Sky
Survey plates. Since the nature of the first object
is known, we performed spectroscopy only for the
second source to determine its nature. Balmer emission
lines corresponding to zero redshift and broad
absorption lines are clearly seen in our spectrum
(Fig. 1c). Such an optical spectrum can correspond
to the spectrum of an M dwarf and a star with an
active chromosphere.

The XRT X-ray spectra of sources 1 and 2 (the
observations were carried out several times in June-October 2010, ObsID 41177, the total exposure time
is $\sim9.3$ ks) show a striking difference (Fig. 1d). The
spectrum of the first of them is typical of quasars
and can be fitted by a power-law dependence of the
photon flux density on energy, $dN/dE\propto E^{-\Gamma}$, with a
photon index $\Gamma=1.7\pm0.4$. There is absorption in the
source's spectrum that exceeds the interstellar one in
this direction ($\sim4\times10^{20}$ cm$^{-2}$; Dickey and Lockman
1990), but the significance of this measurement
is not very high, $N_{H}=0.32^{+0.26}_{-0.16}\times10^{22}$ cm$^{-2}$. The
X-ray spectrum of the second objects turns out to be
considerably softer; no signal is detected above 2 keV
and the spectrum itself corresponds to blackbody radiation
with a temperature of $\sim2\times10^{6}$ K and a flux of
$\simeq10^{-13}$ \flux\ in the $0.5-2$ keV energy band.
Such a temperature is typical of the chromospheres of
late-type stars, including M dwarfs. Thus, this source
(within the BAT error circle) cannot provide the hard
X-ray emission from SWIFT J1553.6+2606 and the
latter is a quasar at redshift $0.1664$ with a luminosity
of $\sim4\times10^{43}$ \ergs\ in the $2-10$ keV energy band.

\bigskip

\subsection*{SWIFT\,J1852.2+8424}

\begin{figure}
\centering
\includegraphics[width=\columnwidth,bb=140 260 547 672,clip]{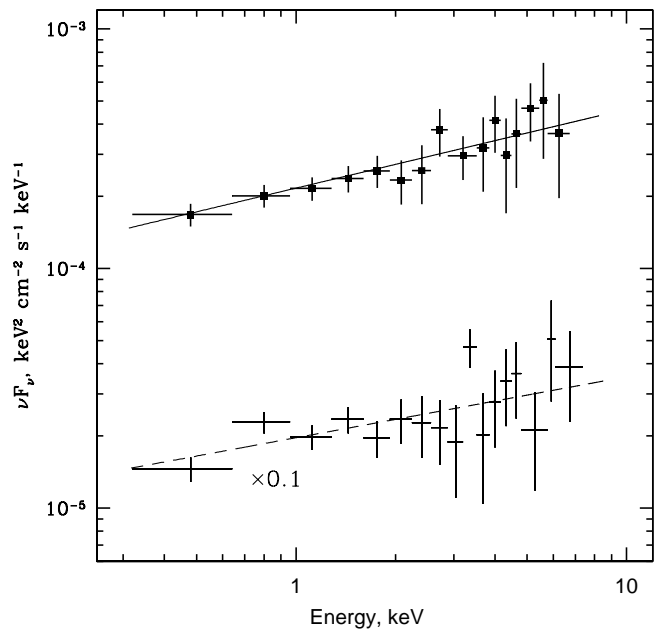}
\caption{XRT energy spectra of sources A (dots) and B
(crosses). The solid and dashed lines, respectively, indicate
the power-law best fits. As in Fig. 1, the spectrum of
source B was multiplied by 0.1 for clarity.  \label{swiftj18522b}}
\end{figure}

SWIFT J1852.2+8424 is another hard X-ray
source from the Swift survey with two possible soft
X-ray counterparts located within its BAT error
circle (Fig. 2a). In contrast to the preceding object,
the intensities of both soft X-ray sources (following
Baumgartner et al. 2013, they are designated by
letters A and B here) are essentially identical in the
$2-10$ keV energy band ($F_{X,A}\simeq7.6\times10^{-13}$ and $F_{X,B}\simeq9.1\times10^{-13}$ \flux, respectively). In
the optical wavelength range, they correspond to
objects with magnitudes $m_{r,A}\simeq17.4$ and $m_{r,B}\simeq15.6$ and the coordinates given in Table 3 (see also Fig. 2b).
The RTT-150 optical spectra also turn out to be
very similar (Figs. 2c and 2d). They exhibit sets
of emission lines typical of Seyfert 1 galaxies --
broad Balmer hydrogen lines, narrow O[III], 4959,
5007 oxygen lines, etc. The redshifts of the galaxies
measured from narrow lines are $z = 0.1828$ and $z =
0.2249$ for sources A and B, respectively.

The XRT X-ray spectra are typical of active
galactic nuclei -- they are well fitted by a simple
power law with photon indices $\Gamma_A=1.75\pm0.07$ and $\Gamma_B=1.67\pm0.07$ (Fig. 3). The luminosities
of the galaxies in the 2-10 keV energy band are
$L_{X,A}\simeq0.7\times10^{44}$ and $L_{X,B}\simeq1.4\times10^{44}$ \ergs, according
to their redshifts. Thus, the X-ray emission from
SWIFT J1852.2+8424 detected at energies $>15$ keV
is the sum of the emissions from two Seyfert 1
galaxies, with the contribution from each galaxy
being approximately the same.

\bigskip

\subsection*{SWIFT\,J1852.8+3002}

\begin{figure*}
\vbox{
\hbox{
\includegraphics[width=\columnwidth,bb=50 164 560 620,clip]{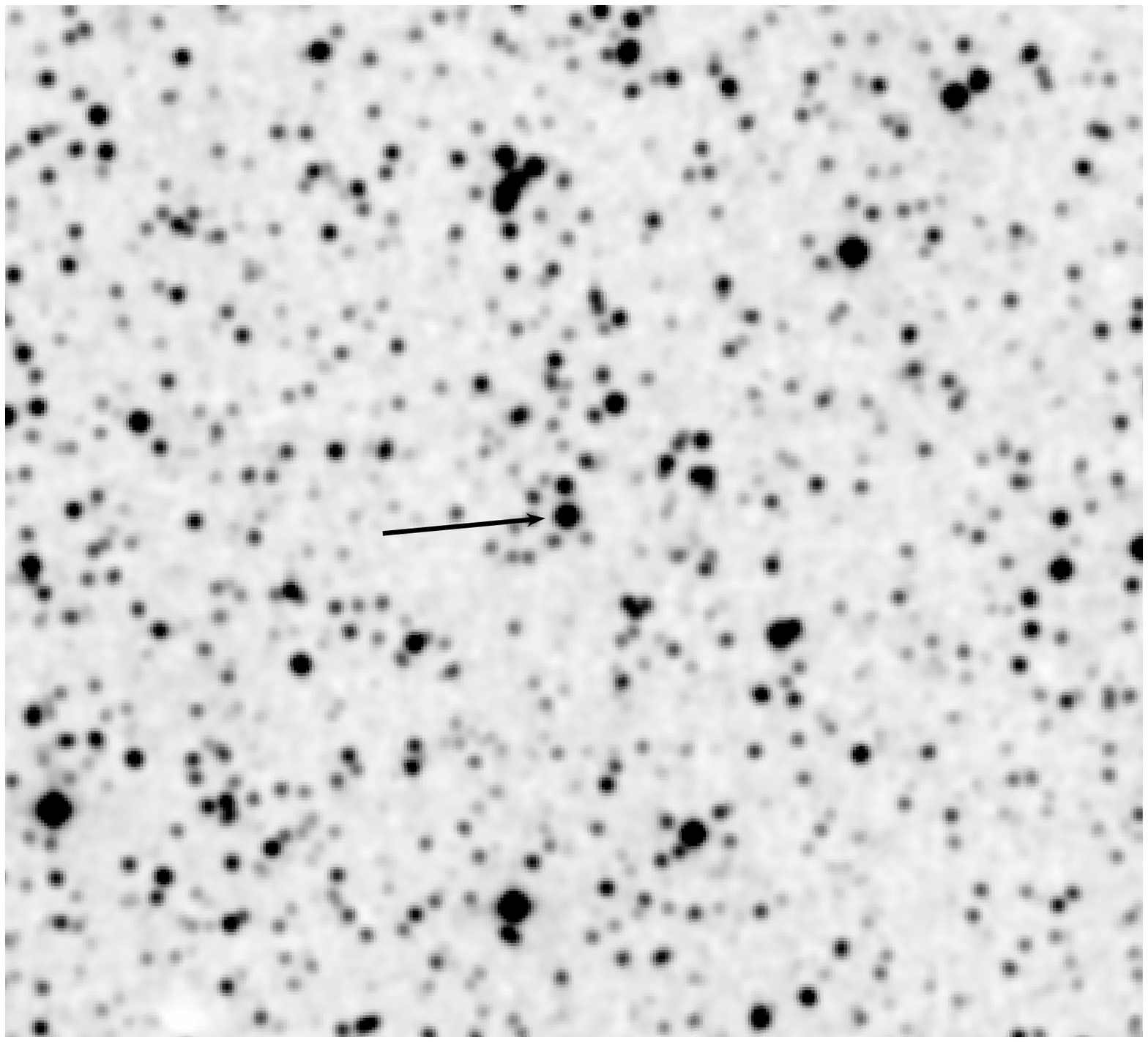}
\hspace{3mm}\includegraphics[width=0.95\columnwidth,bb=29 167 568 672,clip]{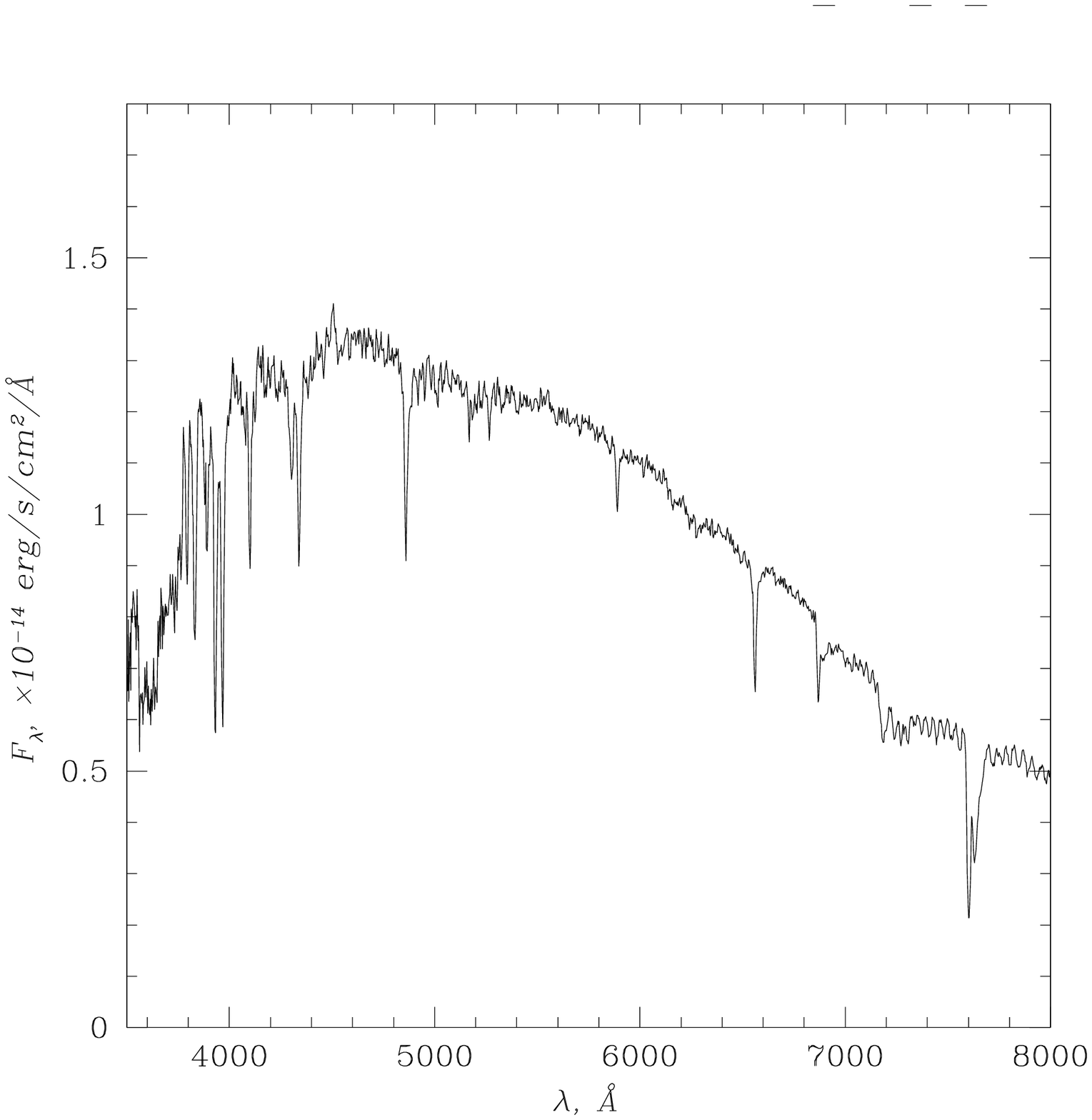}
}
\vspace{3mm}
\hbox{
\includegraphics[width=0.93\columnwidth,bb=137 272 547 672,clip]{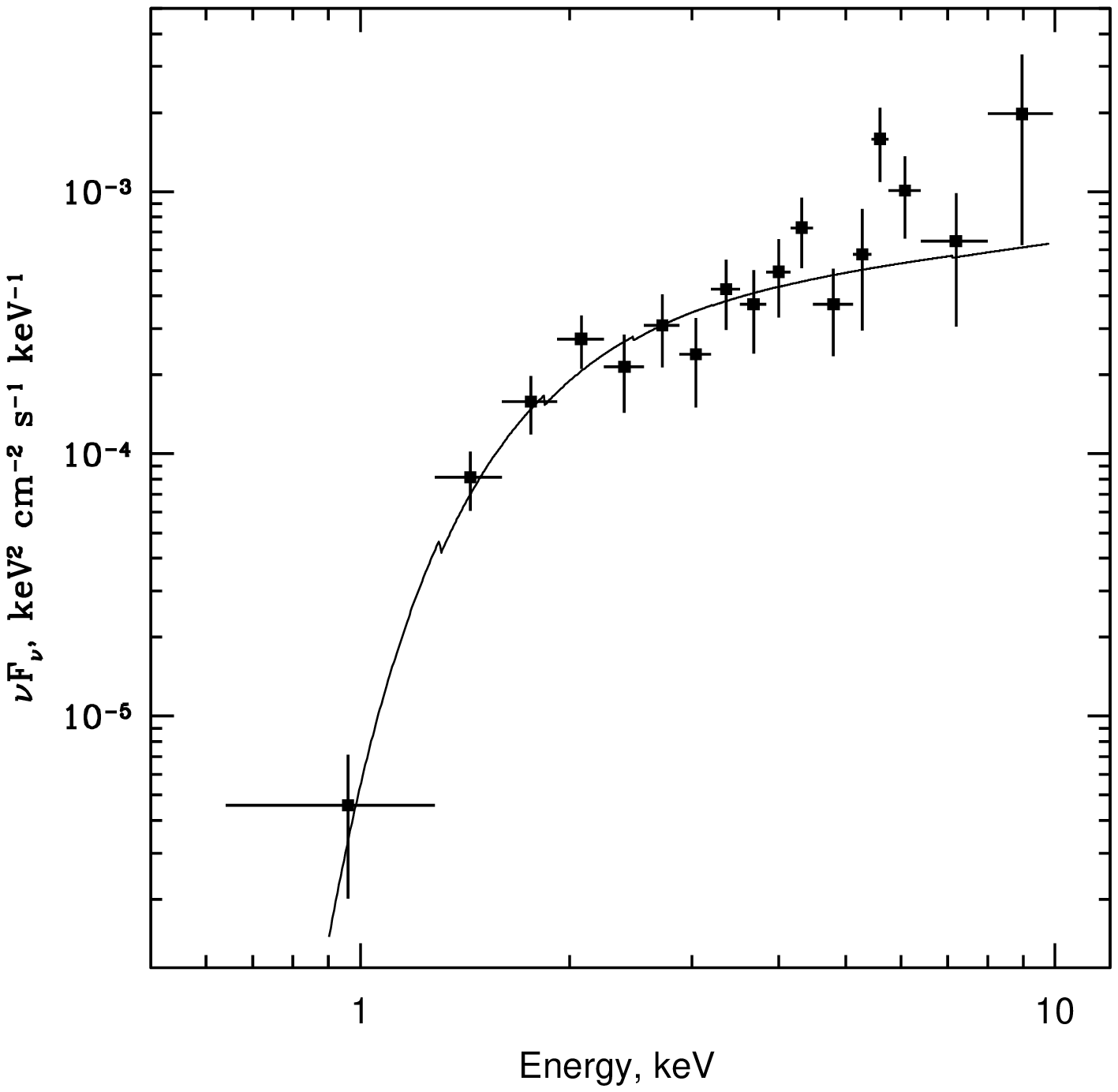}
\includegraphics[width=1.07\columnwidth,bb=37 169 574 621,clip]{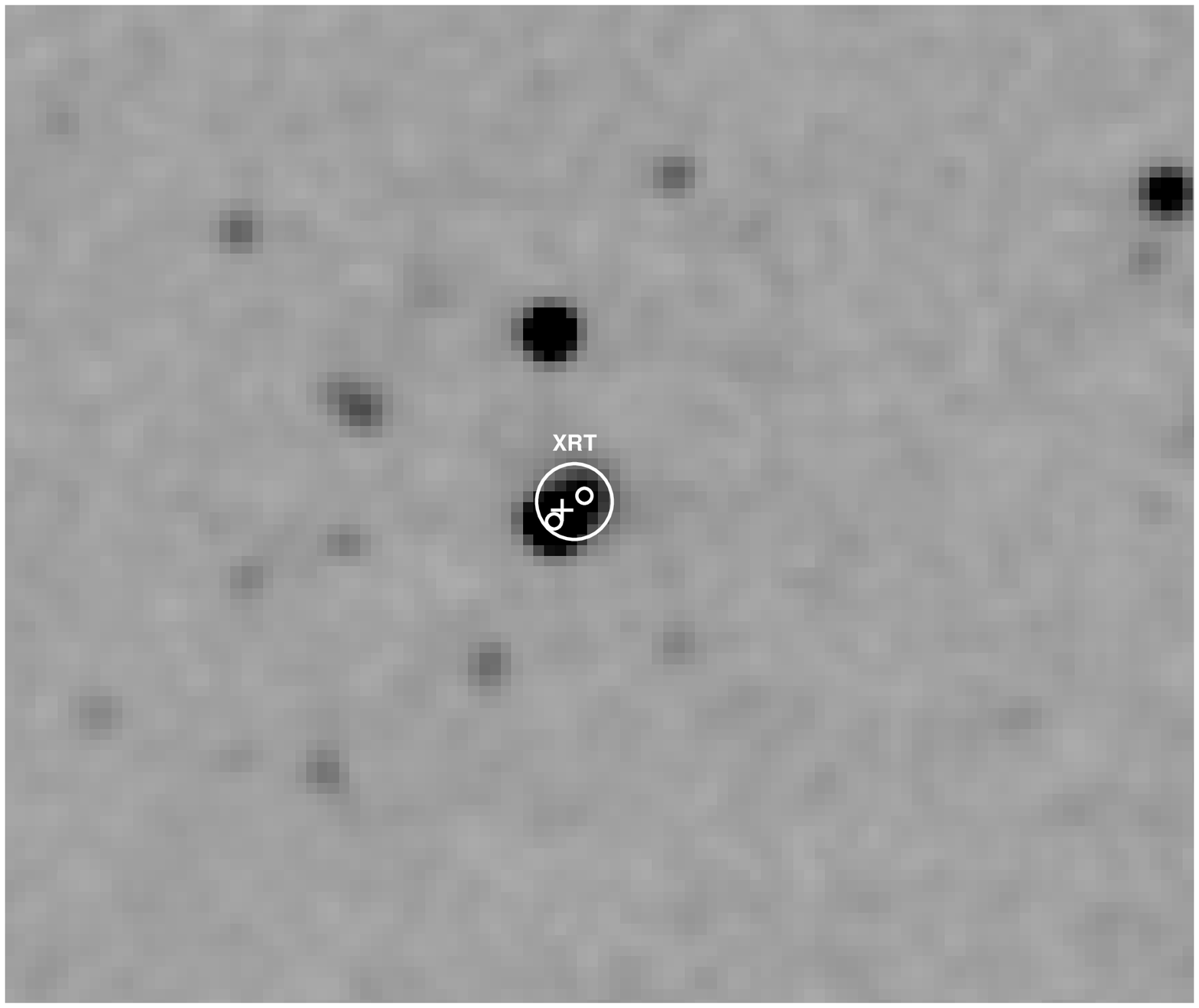}
}
}
\caption{(a) Optical image of the sky around SWIFT J1852.8+3002. The arrow indicates the position of the optical counterpart.
(b) The RTT-150 optical spectrum of the source. (c) The XRT energy spectrum of the source. The solid line indicates the
power-law best fit with absorption at low energies. (d) An enlarged infrared (2MASS, the H band) image of the sky region
around SWIFT J1852.8+3002. The big white circle indicates its XRT error circle, the cross indicates the position of the optical
star according to the USNO-B1 catalog, and the small circles indicate the positions of infrared objects from the 2MASS
catalog.\label{swiftj18528}}
\end{figure*}

According to the Palomar Digital Sky Survey
plates, a fairly bright star with coordinates (J2000)
RA=18$^h$ 52$^m$ 49.590$^s$, Dec=30\deg 04\arcmin 26.48\arcsec\
(Fig. 4a) and a magnitude $m_{r}\simeq12.5$ lies almost at
the center of the XRT error circle for
SWIFT J1852.8+3002 (the observations in May.
June 2010, ObsID 40998, the total exposure time is
$\sim9.6$ ks). A set of absorption lines corresponding to
$H\alpha$, $H\beta$, the CaII H and K doublet, and the Balmer
jump below $\sim3800\AA$ (Fig. 4b) typical of F5 III stars
are clearly seen in its RTT-150 optical spectrum.
Comparison of the apparent and absolute magnitudes
for stars of this type gives an estimate of its distance, $\sim1$kpc.

On the other hand, the X-ray spectrum of
SWIFT J1852.8+3002 fitted by a simple power
law with a slope $\Gamma\simeq1.7$ exhibits significant absorption,
$N_{H}\simeq1.6\times10^{22}$ cm$^{-2}$ (Fig. 4c). This
value is approximately an order of magnitude higher
than the column density of the matter in our Galaxy
in this direction, $N_H\simeq1.4\times10^{21}$ cm$^{-2}$ (Dickey
and Lockman 1990), suggesting the presence of a
substantial amount of matter in the binary itself
possibly associated with the stellar wind from the
optical companion. The X-ray flux from the source
is $\simeq10^{-13}$ \flux\ in the 2-10 keV energy
band, which corresponds to its luminosity $L_X\simeq10^{30}$ \ergs\ for the above estimate of the distance
to the binary.

Such a low luminosity in combination with an
optical F5 III companion is rather unusual for Xray
binaries, especially if the source's hard X-ray
spectrum is taken into account. A study of infrared
maps and catalogs for this sky region showed
that there are actually two close objects with coordinates
RA=18$^h$ 52$^m$ 49.647$^s$, Dec=30\deg 04\arcmin 25.44\arcsec\
and RA=18$^h$ 52$^m$ 49.431$^s$, Dec=30\deg 04\arcmin
27.80\arcsec\ and
JHK (2MASS) magnitudes $\simeq12.25, 12.08, 11.82$ and $\simeq12.71, 13.75, 12.11$, respectively, at the position
of the bright optical star (Fig. 4d). The first of these
objects probably corresponds to the optical star from
the Palomar Digital Sky Survey whose spectrum
was taken with RTT-150, while the second object
whose position is closer to the center of the error
circle for the X-ray source SWIFT J1852.8+3002
(Table 1) is its optical counterpart. However, further
studies, in particular, infrared spectroscopy, are
needed to ultimately answer this question and to
determine the object's class. In conclusion, note
that SWIFT J1852.8+3002 lies fairly high above the
Galactic plane ($b\simeq13$\deg), which is atypical of high mass
X-ray binaries whose vertical distribution does
not exceed a hundred parsecs (see, e.g., Lutovinov
et al. 2013).

\bigskip

\subsection*{IGR\,J22534+6243}

\begin{figure*}
\vbox{
\hbox{
\includegraphics[width=1.02\columnwidth,bb=50 164 560 625,clip]{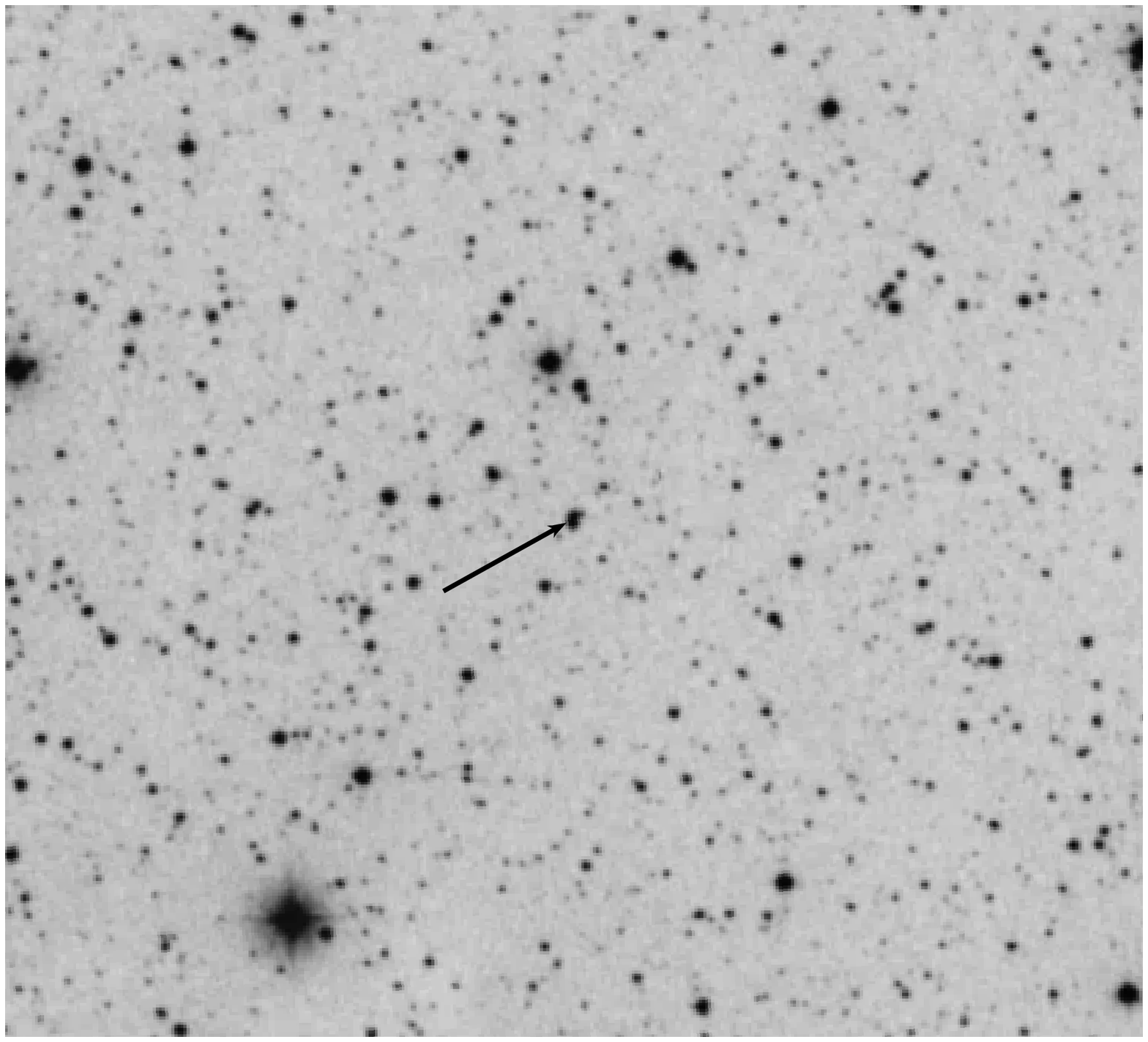}
\hspace{3mm}\includegraphics[width=0.98\columnwidth,bb=33 167 568 681,clip]{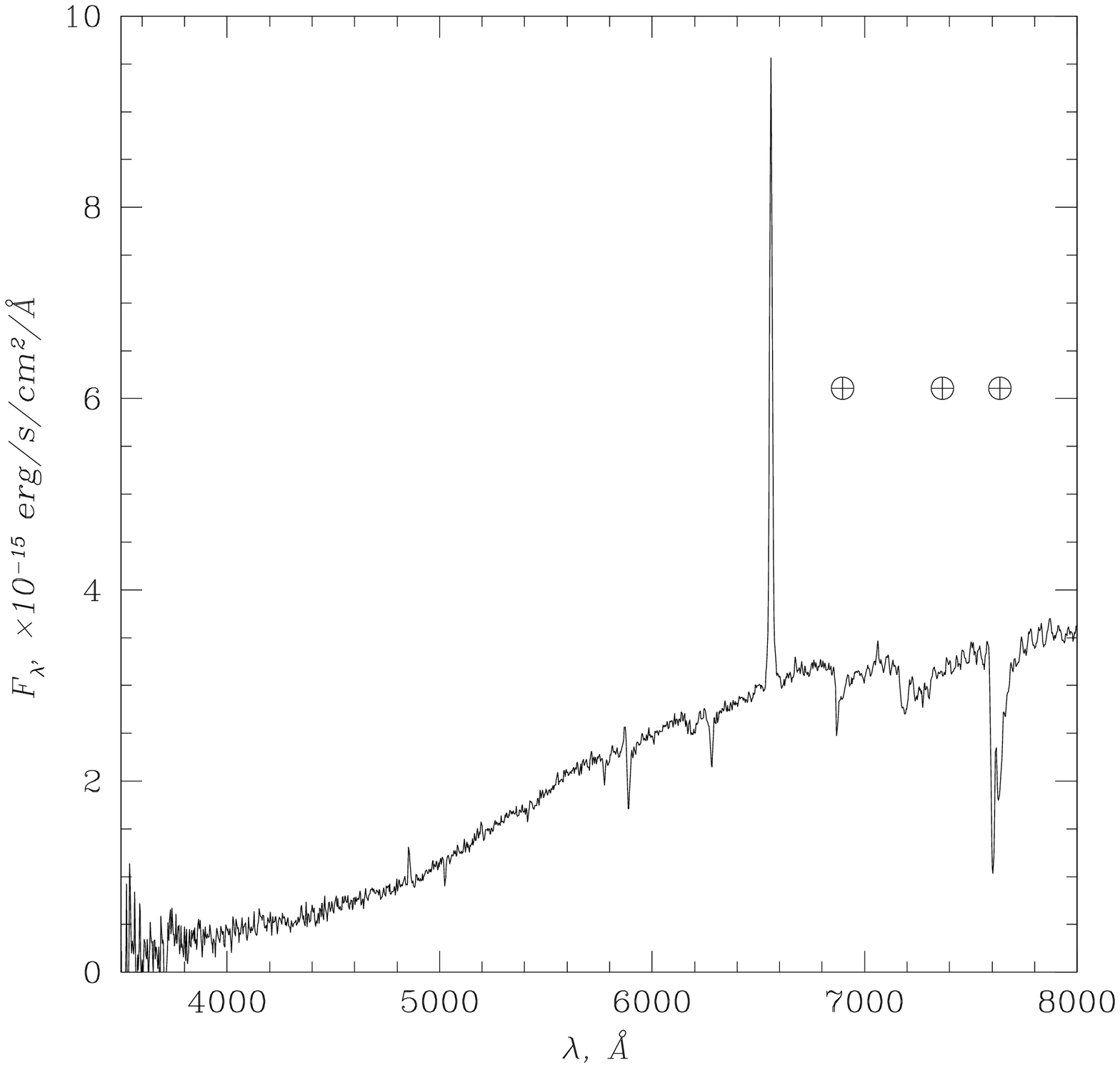}
}

\vspace{3mm}
\hbox{
\includegraphics[width=1.015\columnwidth,bb=45 184 583 692,clip]{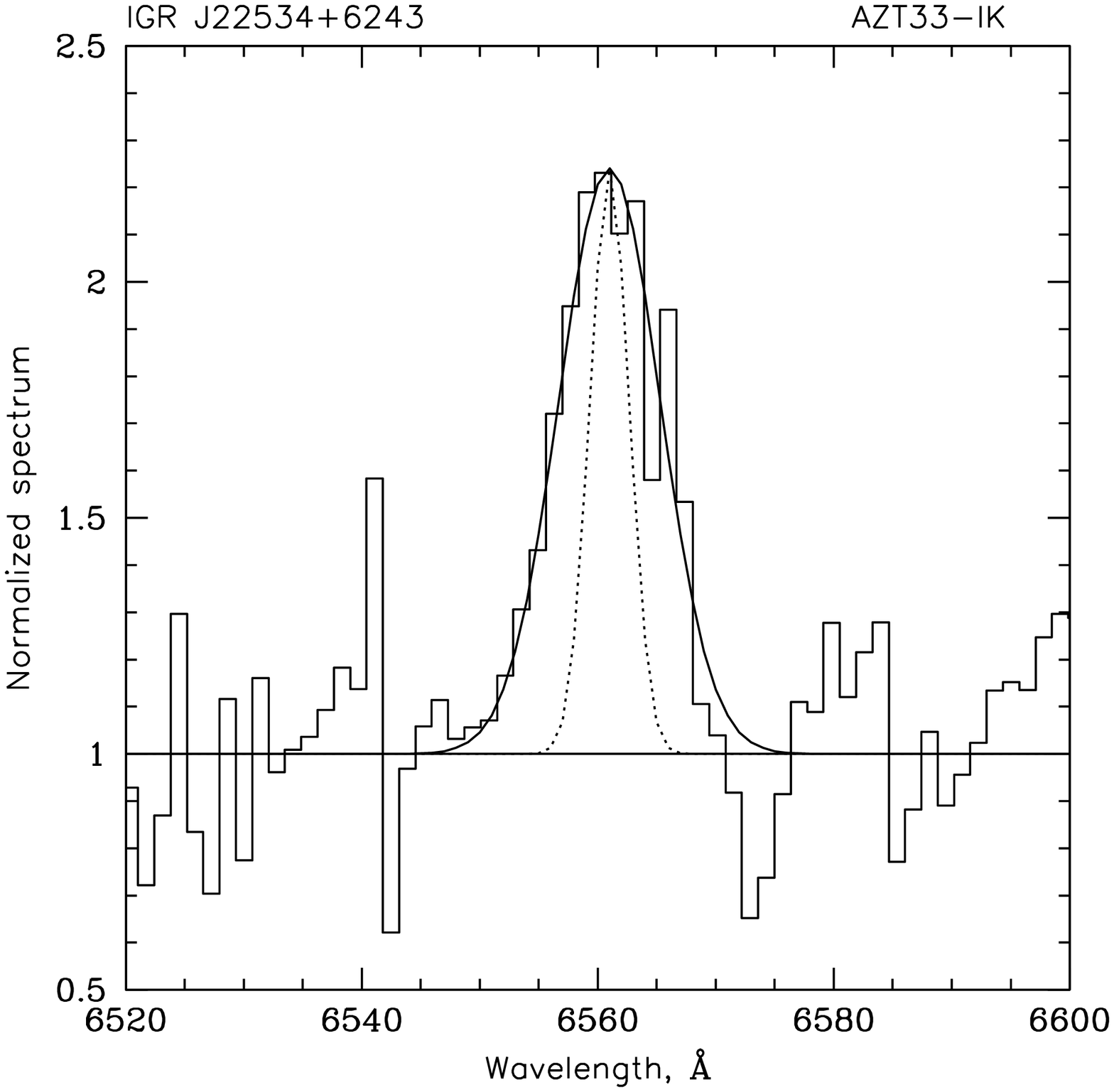}
\hspace{3mm}\includegraphics[width=0.98\columnwidth,bb=138 272 547 672,clip]{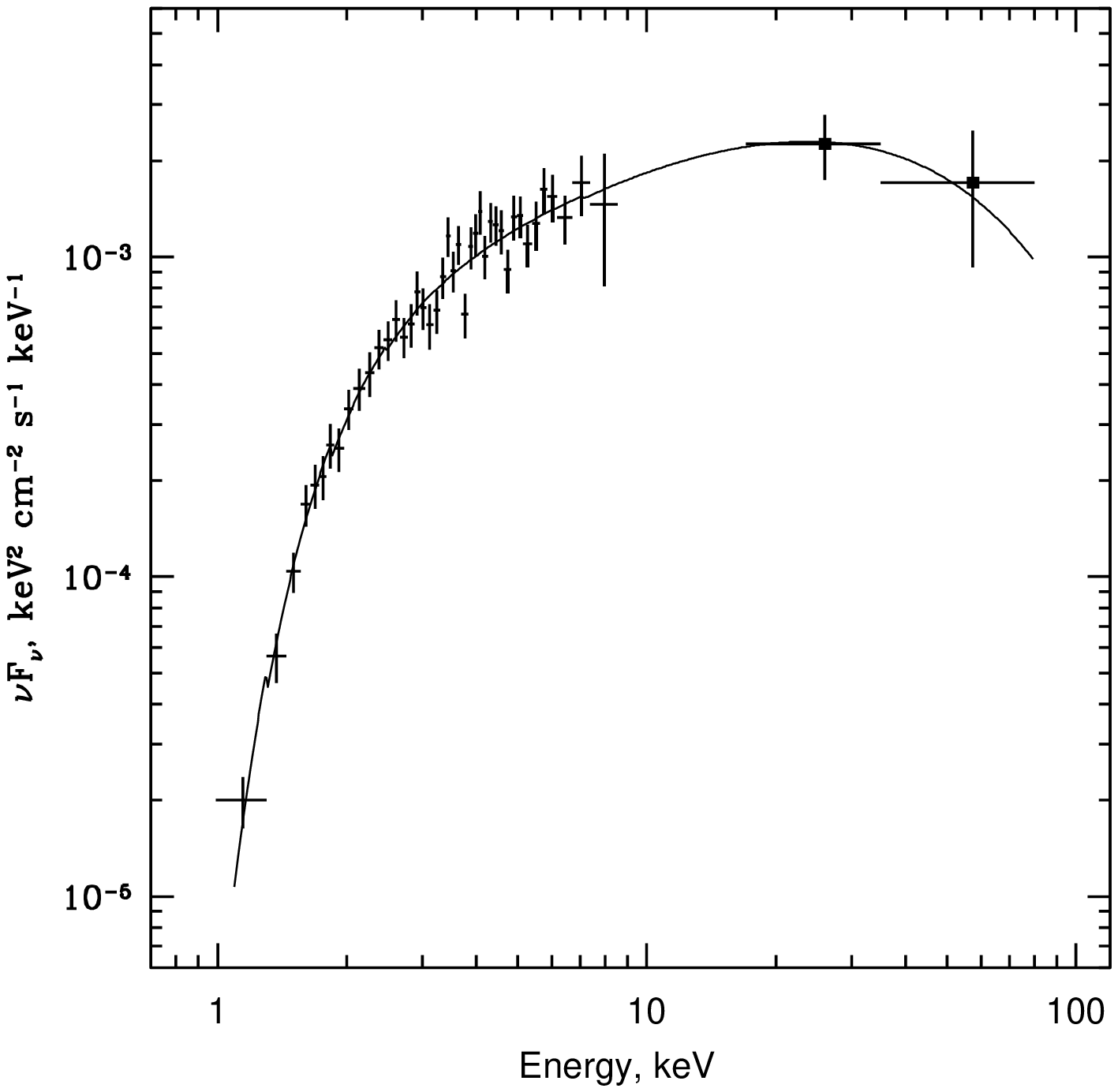}
}
}
\caption{(a) Optical image of the sky around IGR J22534+6243. The arrow indicates the position of the optical counterpart. (b)
The RTT-150 optical spectrum of the source taken in July 2012. (c) Part of the AZT-33IK optical spectrum for the source near
$H\alpha$ obtained in March 2013. The solid line indicates the best fit to the line by a Gaussian profile; the dotted line indicates the
characteristic profile of atmospheric lines obtained by assuming $\Delta \lambda/\lambda = {\rm const}$. (d) The source's broadband energy spectrum
from Chandra (crosses, ObsID. 10811) and INTEGRAL (dots) data. The solid line indicates the best fit by a simple power law
with low-energy absorption and a high-energy cutoff. \label{igrj22534}}
\end{figure*}

The hard X-ray emission from IGR J22534+6243
was detected only on the total map of the Galactic
plane constructed from nine-year-long INTEGRAL
observations (Krivonos et al. 2012). No significant
variations of the source's intensity (in particular, outbursts)
were detected on its reconstructed light curve
for 2003--2012 and the mean flux was $F_X=(0.6\pm0.1)\times10^{-11}$ \flux\ in the $17-60$ keV energy
band.

As has already been said above, the study of the
archival data showed that the sky region around
IGR J22534+6243 previously fell within the
XRT/Swift field of view when the afterglow from
GRB 060421 was investigated (ObsId. 00206257,
April 21--24, 2006, the total exposure time is $\sim72$ ks)
and the Chandra field of view in March--April 2009
(ObsIDs. 9919, 9920, 10810, 10811, 10812; the
exposure time of each pointing was $23-28$ ks). It
should be noted that during the Chandra observations
IGR J22534+6243 was almost at the edge of the
telescope's field of view, but it was detected at a
statistically significant level in all pointings. The
source's coordinates measured from these observations
closely coincided with those measured from
XRT data (see Table 1), which allowed the optical
counterpart of IGR J22534+6243 to be determined.
It turned out to be a fairly bright object with coordinates
(J2000) RA=22$^h$ 53$^m$
55.130$^s$, Dec=62\deg 43\arcmin 36.90\arcsec\ (Fig. 5a) and a magnitude $m_{r}\simeq13.0$,
which is clearly seen in both optical and infrared
(2MASS J22535512+6243368) wavelength ranges.

The RTT-150 optical spectrum of the object
corresponds to a star of early spectral type O-B
(Fig. 5b). At the same time, an intense Balmer $H\alpha$
emission line and a weaker $H\beta$ line are clearly seen
in this spectrum. The spectrum
near the emission $H\alpha$ line was taken at the AZT-33IK
telescope with a spectral resolution better than that
of RTT-150 (Fig. 5c). The equivalent width of the
emission line is $13\pm1\AA$. The $H\alpha$ emission lines with
such equivalent widths are commonly observed from
Be disks (see, e.g., Clark et al. 2001). Since the AZT-33IK spectrum has a good resolution, we managed
to detect a finite width of the $H\alpha$ emission line.
At an instrumental resolution of our spectroscopic
data of about $3.8\AA$ (FWHM), the observed $H\alpha$ line width
was about $4.3\AA$. Thus, we may conclude that in
the source itself the line is broadened with typical
velocities of about 180 km s$^{-1}$, which is also a
commonly observed characteristic of the emission
lines in Be systems, and is associated with a rotating
equatorial disk around a Be star. It should also be
noted that the line equivalent width changed by more
than a factor of 2 in half a year elapsed between the
RTT-150 and AZR-33IK observations (it was $\simeq33\AA$
in July 2012), which may be indicative of equatorial disk
evolution.

Significant detection of the source by different
instruments in different time intervals allowed us
not only to carry out spectral and timing analysis
of its emission but also to trace the evolution of
its parameters. In particular, the source's intensity
in the $2-10$ keV energy band remained essentially
constant during the Swift and Chandra observations
at $F_X\simeq(2.5-2.9)\times10^{-12}$ \flux,
increasing only once (ObsID. 10810) to $F_X\simeq3.1\times10^{-12}$ \flux. However, given the typical flux
measurement error of $\simeq(0.13-0.16)\times10^{-12}$ \flux, such a change may be considered
insignificant. The same can also be said
about the source's spectrum, which can be well fitted
by a simple power law with low-energy absorption.
The slope of the spectrum varies insignificantly between
$\Gamma=1.35\pm0.14$ and $\Gamma=1.63\pm0.10$, becoming
slightly harder, $\Gamma=1.18\pm0.13$, at the beginning
of the series of Chandra observations (ObsID. 9920,
April 16, 2009). We found no correlations between
the variations of the flux from the source and the
hardness of its spectrum. The parameters of the
spectrum for IGR J22534+6243 determined using
data from the ROSAT observatory that observed this
sky region on June 18-19, 1993, (ObsID. 500321,
the total exposure time is $\sim18.5$ ks) agree with the
Swift and Chandra measurements, but the typical
measurement errors turned out to be considerably
larger than those given above.

\begin{table}[]
\centering
\caption{ }
\vspace{1mm}
  \begin{tabular}{ll}
    \hline
    \vspace{0.5mm}
    Data, MJD & Period, s \\
    \hline
    \vspace{0.5mm}
    49156.08 & $46.4040\pm0.0008$ \\
    53846.81 & $46.6148\pm0.0002$ \\
    54937.45 & $46.6799\pm0.0031$ \\
    54949.29 & $46.6695\pm0.0032$ \\
    54954.07 & $46.6718\pm0.0024$ \\
    54958.85 & $46.6723\pm0.0042$ \\
    54959.14 & $46.6658\pm0.0033$ \\[1mm]
    \hline
  \end{tabular}
\end{table}

The measured absorption, $N_H=(2.08-2.27)\times10^{22}$ cm$^{-2}$, slightly exceeds the column density
of matter in our Galaxy in this direction, $N_H\simeq10^{22}$ cm$^{-2}$ (Dickey and Lockman 1990). This suggest
the presence of an additional amount of material
in the binary itself possibly associated with the stellar
wind from the optical companion.
Taking into account the stability of the parameters
for the X-ray spectrum of IGR J22534+6243 over a
long time, we may extend it to the hard energy region
using the INTEGRAL observational data. Such a
broadband spectrum is shown in Fig.5c. Since the
source is weak, a relatively significant signal from
it can be registered only in broad channels. However,
these measurements suggest a cutoff in the
spectrum at high energies. The characteristic cutoff
energy, $E_{cut}\simeq25-30$ keV, turns out to be slightly
higher than that commonly observed in high-mass Xray
binaries with neutron stars (see, e.g., Filippova
et al. 2005). However, it should be noted that such
a hard spectrum was recorded recently from low luminosity
X-ray pulsars that are members of binaries
with Be stars (see, e.g., Tsygankov et al. 2012;
Lutovinov et al. 2012c).

\begin{figure}
\centering
\includegraphics[width=\columnwidth,bb=73 270 495 690]{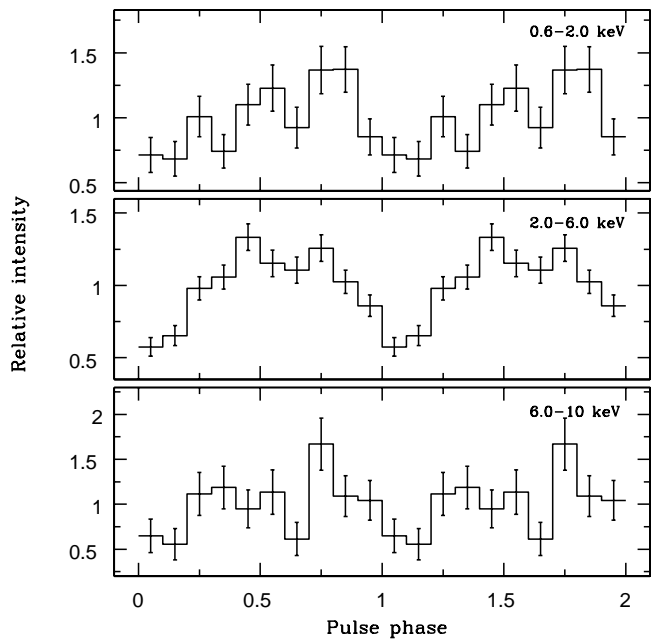}
\caption{Pulse profile for the X-ray pulsar
IGR J22534+6243 in various energy bands obtained
from one of the Chandra observations (ObsID.10811)
and folded with the corresponding period.   \label{pprofile}}
\end{figure}

\begin{table*}[]
\centering
  \caption{Optical identification of the hard X-ray sources}
  \label{tab:2}
  \vspace{1mm}
  \begin{tabular}{lccll}
    \hline
    Name & RA & Dec & Type & $z$ \\
    \hline
    \hline
     SWIFT\,J1553.6+2606  &  15$^h$ 53$^m$ 34.734$^s$ &  26\deg 14\arcmin 41.45\arcsec & QSO      & 0.1664  \\
     SWIFT\,J1852.2+8424A &  18$^h$ 50$^m$ 25.090$^s$ &  84\deg 22\arcmin 44.69\arcsec & Sy1      & 0.1828  \\
     SWIFT\,J1852.2+8424B &  18$^h$ 46$^m$ 49.689$^s$ &  84\deg 25\arcmin 05.58\arcsec & Sy1      & 0.2489  \\
     SWIFT\,J1852.8+3002  &  18$^h$ 52$^m$ 49.431$^s$ &  30\deg 04\arcmin 27.80\arcsec & XRB/HMXB(?) &   \\
     IGR\,J22534+6243     &  22$^h$ 53$^m$ 55.130$^s$ &  62\deg 43\arcmin 36.90\arcsec & HMXB, X-ray pulsar  &   \\
     \hline

    \hline
  \end{tabular}
\end{table*}

Swift and Chandra data allowed Halpern (2012)
to detect X-ray pulsations with a period of $\sim46.7$ s
in the light curve for IGR J22534+6243, suggesting
the presence of a neutron star as a compact object in
the binary. Pulsations with a similar period were also
found in the ROSAT archival data for this sky region
(Israel and Rodriguez 2012). We analyzed in detail
all observational data and traced the evolution of the
source's pulsation period, which is presented in Table 2
(the first measurement is based on ROSAT data,
the second measurement is based on XRT data, and
the remaining measurements are based on Chandra
data).

It can be seen from the Table 2 that in the 16-year
time interval between the ROSAT and Chandra observations,
the neutron star spun down significantly,
with the mean spin down rate having remained almost
the same over the entire period of ROSAT, Swift, and
Chandra observations near $\dot P/P\simeq3.5\times10^{-4}$ yr$^{-1}$,
which is typical of X-ray pulsars (see, e.g., Lutovinov et al. 1994; Bildsten et al. 1997). At the same time,
during almost a month of Chandra observations, the
pulsation period changed insignificantly, remaining
mainly near 46.674 s within the measurement errors.
The measurement errors themselves were determined
by the so-called bootstrap method (for more detail,
see Lutovinov et al. 2012c).

One of the most important characteristics for
an X-ray pulsar is its pulse profile. As a rule, the
pulse shape remains fairly stable for each specific
pulsar over a long time, although it can depend on
its luminosity and energy (see, e.g., Lutovinov and
Tsygankov 2009). Our study of the pulse profile
for IGR J22534+6243 showed that in all Chandra
and Swift observations it has a similar shape (see
Fig. 6) --- a complex double-peak, triple-peak structure
is seen in the softest $0.6-2.0$ keV energy band,
which turns into one broad peak in the $2-6$ keV
energy band, with the signatures of this structure
remaining in it. As the energy increases, several
peaks again begin to manifest themselves in the
pulse profile. The latter may be due to a shortage
of statistics at energies $>6$ keV, while the multi-peak
structure at soft energies can be a consequence of
absorption in the binary. The pulsed fraction in the
$2-6$ keV energy band is $\sim40$\%.

\section{Conclusions}
\label{sec:concl}

We made an optical identification of four Xray
sources from the INTEGRAL and Swift catalogs
and determined the nature of three of them.
Two of these sources are extragalactic in nature:
SWIFT J1553.6+2606 is a quasar at redshift $z\simeq0.1164$; the detected flux from SWIFT J1852.2+8424
is the sum of the fluxes from two Seyfert 1 galaxies
of approximately the same intensity at redshifts $z\simeq0.1828$ and $z\simeq0.2489$. The other two objects are
located in our Galaxy: IGR J22534+6243 belongs to
the class of X-ray pulsars (with a pulsation period
of $\simeq46.674$ s) in high-mass X-ray binaries most
likely with a Be companion; SWIFT J1852.8+3002
may also be a high-mass X-ray binary, but infrared
spectroscopy is needed for the ultimate answer to the
question about its nature.

Obtained results are summarized in Table 3, which
gives the coordinates of the optical counterparts of the
X-ray sources, their types, and (for the extragalactic
objects) redshifts.

\bigskip
~\bigskip

\vspace{10mm}

\acknowledgements

This work was supported by the Russian Foundation for Basic Research (project nos. 12-02-01265, 11-02-01328), the Presidium of Russian Academy of Sciences (programs P-21 and OFN-17), the Programs of the President of Russia for
Support of Leading Scientific Schools (project NSh-5603.2012.2) and the Ministry of Education and
Science (Contracts N8701 and N8629). We wish to
thank the TUBITAK National Observatory (TUG,
Turkey), the Space Research Institute of the Russian
Academy of Sciences, and the Kazan State
University for support in using the 1.5-m Russian-Turkish (RTT-150) telescope. We are also grateful to
E.M. Churazov, who developed the IBIS/INTEGRAL
data analysis methods and provided the software.

\parindent=0mm

1. W. Baumgartner, J. Tueller, C. Markwardt, et
al., Astrophys. J. Suppl. Ser. (2013, in press);
arXiv:1212.3336.

2. I. Bikmaev, M. Revnivtsev, R. Burenin, and R. Sunyaev,
Astron. Lett. 32, 588 (2006).

3. I. Bikmaev, R. Burenin, M. Revnivtsev, et al., Astron.
Lett. 34, 653 (2008).

4. L. Bildsten, D. Chakrabarty, J. Chiu, et al., Astrophys. J. Suppl. Ser. 113, 367 (1997).

5. A. Bird, A. Bazzano, L. Bazzani, et al., Astrophys.
J. Suppl. Ser. 186, 1 (2010).

6. R. Burenin, A. Meshcheryakov, M. Revnivtsev, et al.,
Astron. Lett. 34, 367 (2008).

7. R. Burenin, I. Bikmaev, M. Revnivtsev, et al., Astron.
Lett. 35, 71 (2009).

8. J. S. Clark, A. E. Tarasov, A. T. Okazaki, et al.,
Astron. Astrophys. 380, 615 (2001).

9. G. Cusumano, V. La Parola, A. Segreto et al., Astron.
Astrophys. 524, 64 (2010).

10. J. Dickey and F. Lockman, Ann. Rev. Astron. Astrophys.
28, 215 (1990).

11. E. Filippova, S. Tsygankov, A. Lutovinov, and
R. Sunyaev, Astron. Lett. 31, 729 (2005).

12. N. Gehrels, G. Chinkarini, P. Giommi, et al., Astrophys.
J. 611, 1005 (2004).

13. J. Halpern, Astron. Telegram 4240, 1 (2012).

14. G. Israel and G. Rodriguez, Astron. Telegram4241, 1 (2012).

15. D. I. Karasev, A. A. Lutovinov, M. G. Revnivtsev, and
R. A. Krivonos, Astron. Lett. 38, 629 (2012).

16. R. Krivonos, S. Tsygankov, M. Revnivtsev, et al.,
Astron. Astrophys. 523, A61 (2010a).

17. R. Krivonos, M. Revnivtsev, S. Tsygankov, et al.,
Astron. Astrophys. 523, A107 (2010b).

18. R. Krivonos, S. Tsygankov, A. Lutovinov, et al., Astron.
Astrophys. 545, A27 (2012).

19. R. Landi, L. Bassani, N. Masetti, et al., Astron. Telegram
4166, 1 (2012).

20. A. Lutovinov, S.Grebenev, R.Sunyaev and M.Pavlinsky, Astron. Lett. 20, 538 (1994)

21. A. Lutovinov and S. Tsygankov, Astron. Lett. 35, 433
(2009).

22. A. Lutovinov, R. Burenin, M. Revnivtsev, and I. Bikmaev,
Astron. Lett. 38, 1 (2012a).

23. A. Lutovinov, R. Burenin, M. Revnivtsev, et al., Astron.
Lett. 38, 281 (2012b).

24. A. Lutovinov, S. Tsygankov, and M. Chernyakova,
Mon. Not. R. Astron. Soc. 423, 1978 (2012c).

25. A. Lutovinov, M. Revnivtsev, S. Tsygankov, and
R. Krivonos, Mon. Not. R. Astron. Soc. 431, 327 (2013); arXiv:1302.0728.

26. N. Masetti, R. Landi, M. Pretorius, et al., Astron.
Astrophys. 470, 331 (2007).

27. N. Masetti, P. Parisi, E. Palazzi, et al., Astron. Astrophys.
519, 96 (2010).

28. N. Masetti, E. Jimenes-Bailon, V. Chavushyan et al.,
Astron. Telegram 4248, 1 (2012).

29. P. Parisi, N. Masetti, A. Rojas, et al., in Proceedings
of the 9th INTEGRAL Workshop on an INTEGRAL
View of the High-Energy Sky (the first 10 years),
arXiv:1302.6117.

30. J. Tomsick, S. Chaty, J. Rodriguez, et al., Astrophys.
J. 685, 1143 (2008).

31. J. Tomsick, S. Chaty, J. Rodriguez, et al., Astrophys.
J. 701, 811 (2009).

32. S. Tsygankov, R. Krivonos, and A. Lutovinov, Mon.
Not. R. Astron. Soc. 421, 2407 (2012).

33. J. Tueller, W. Baumgartner, C. Markwardt, et al.,
Astrophys. J. Suppl. Ser. 186, 378 (2010).

34. C. Winkler, T. Courvoisier, G. Di Cocco, et al., Astron.
Astrophys. 411, L1 (2003).

\end{document}